\documentclass[conference]{IEEEtran}
\IEEEoverridecommandlockouts
\usepackage{cite}
\usepackage{amsmath,amssymb,amsfonts}
\usepackage{algorithmic}
\usepackage{graphicx}
\usepackage{textcomp}
\usepackage{xcolor}
\usepackage[utf8]{inputenc}
\usepackage{orcidlink}
\usepackage{amsmath,amsfonts}
\usepackage{algorithm}
\usepackage{array}
\usepackage[caption=false,font=normalsize,labelfont=sf,textfont=sf]{subfig}
\usepackage{stfloats}
\usepackage{url}
\usepackage{verbatim}
\usepackage{afterpage}
\usepackage{float}
\usepackage{multirow}
\usepackage{makecell}
\usepackage{array}
\def\BibTeX{{\rm B\kern-.05em{\sc i\kern-.025em b}\kern-.08em
    T\kern-.1667em\lower.7ex\hbox{E}\kern-.125emX}}
\begin{document}

\title{Dual-Precision Floating Point Multiply-Accumulate Processing Engine for AI Applications
\thanks{This work was supported by the Special Manpower Development Program for Chip to Start-Up (SMDP-C2S), Ministry of Electronics and Information Technology (MeitY), Government of India, Grant: EE-9/2/21 - R\&D-E.\\ Corresponding Author: Santosh Kumar Vishvakarma \\Email: skvishvakarma@iiti.ac.in}
}

\author{
    \IEEEauthorblockN{{Shubham Kumar \IEEEauthorrefmark{2}\orcidlink{0009-0006-9997-7189}},
    Vijay Pratap Sharma\IEEEauthorrefmark{1}\orcidlink{0009-0008-1838-5179}, \\  Vaibhav Neema \IEEEauthorrefmark{2}\orcidlink{0000-0003-0922-373X},   Santosh Kumar Vishvakarma\IEEEauthorrefmark{1}\orcidlink{0000-0003-4223-0077}, Senior Member, IEEE.}
    \IEEEauthorblockA{\IEEEauthorrefmark{2}Institute of Engineering and Technology, Devi Ahilya University, Indore, India}
    \IEEEauthorblockA{\IEEEauthorrefmark{1}NSDCS Research Group, Indian Institute of Technology Indore, India}
}

\maketitle
\begin{abstract}

The rapid adoption of low-precision arithmetic in artificial intelligence and edge computing has created a strong demand for energy-efficient and flexible floating-point multiply--accumulate (MAC) units. This paper presents a dual-precision floating-point MAC processing element supporting FP8 (E4M3, E5M2) and FP4 (2 x E2M1, 2 x E1M2) formats, specifically optimized for low-power and high-throughput AI workloads. The proposed architecture employs a novel bit-partitioning technique that enables a single 4-bit unit multiplier to operate either as a standard $4 \times 4$ multiplier for FP8 or as two parallel $2 \times 2$ multipliers for 2-bit operands, achieving the maximum  hardware utilization without duplicating logic. Implemented in 28nm technology, the proposed PE achieves an operating frequency of 1.94~GHz with an area of 0.00396~mm$^{2}$ and power consumption of 2.13~mW, resulting in up to 60.4\% area reduction and 86.6\% power savings compared to state-of-the-art designs, making it well suited for energy-constrained AI inference and mixed-precision computing applications when deployed within larger accelerator architectures.

\end{abstract}

\begin{IEEEkeywords}
Floating-point, Multiply Accumulate, Artificial Intelligence, Processing Element, Hardware Acceleration. 
\end{IEEEkeywords}

\section{Introduction}

\begin{figure}[!t]
\centering
\includegraphics[width=3.45in, height=3.5in]{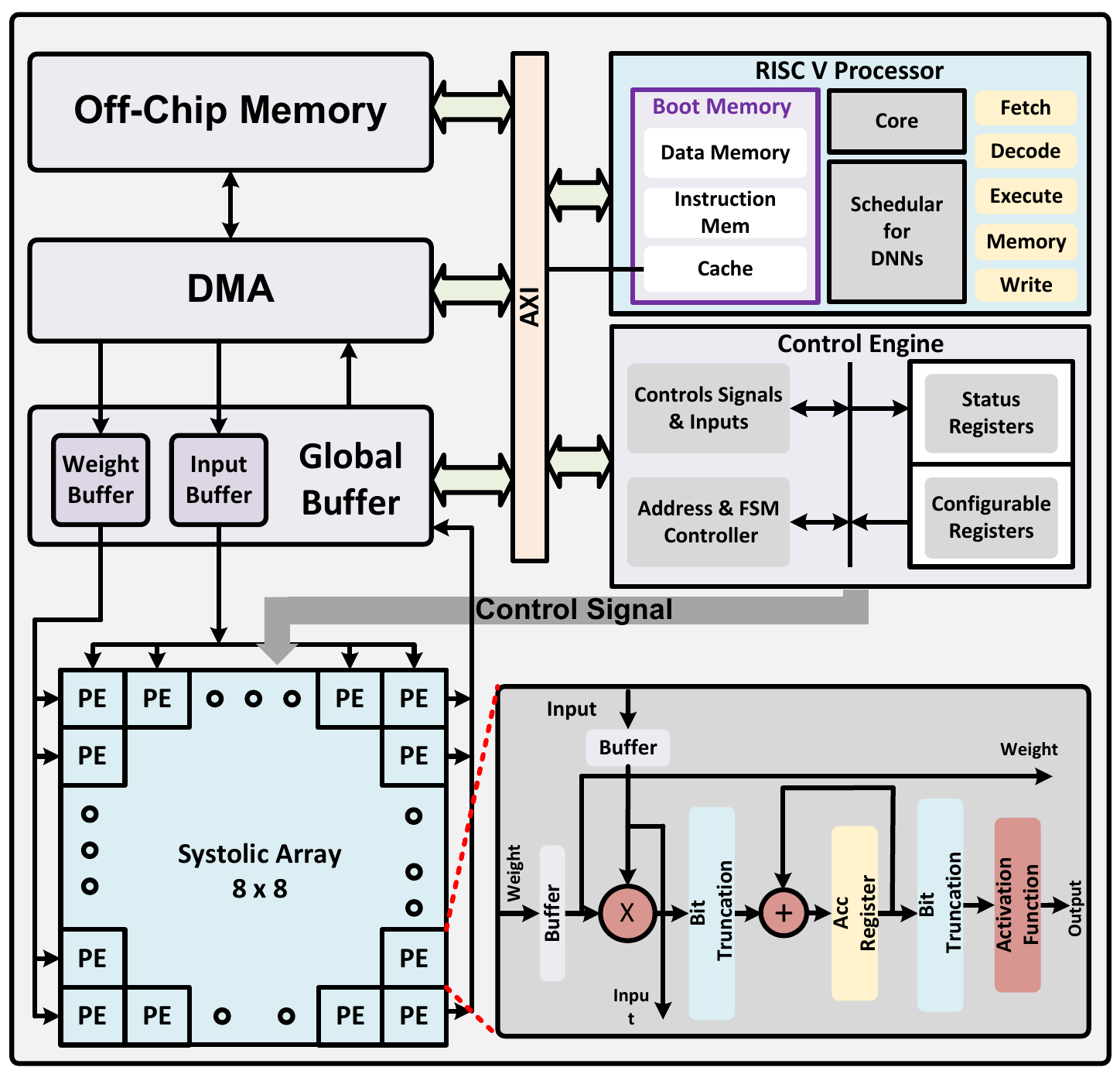}
\caption{Typical AI Accelerator architecture with emphasis on the Processing Element (PE).}
\label{fig:typical_Architecture}
\end{figure}

\IEEEPARstart{T}{he} rapid progress in artificial intelligence (AI) and deep learning has increased the demand for energy-efficient hardware accelerators capable of performing large-scale multiply--accumulate (MAC) operations. Although high-precision formats such as FP16 and FP32 are commonly employed during the training phase, there is a growing trend toward adopting lower-precision formats, including FP8 and FP4, for inference and edge-AI applications due to their reduced memory requirements and improved computational efficiency. Nevertheless, most existing MAC or processing element (PE) architectures are optimized for medium-to-high precision data ($\geq$8~bits) and do not fully exploit the unique data characteristics of FP8 and FP4.
Several adaptable multiple-precision floating-point processing elements (PEs) have been proposed; Fig.~\ref{fig:typical_Architecture} illustrates a typical AI hardware accelerator architecture employing adaptable multiple-precision floating-point processing elements (PEs).
Prior works~\cite{ref1},\cite{ref2,ref3} rely on high-precision split (HPS) paradigms that jointly 
utilize high-precision (HP) and single-precision (SP) formats for training-centric workloads. Similarly, the reconfigurable PEs in~\cite{ref4,ref5}, along with fused multiply--add (FMA) architectures in~\cite{ref6,ref7,ref8,ref9}, employ HPS and low-precision split (LPS) computation schemes, primarily targeting training efficiency rather than inference-optimized execution. Additionally, numerous earlier studies \cite{ref10,ref11,ref12,ref13} utilize quantization methods to facilitate low-precision inference. While these methods are efficient, they frequently necessitate calibration, scaling, and supplementary control mechanisms, which can complicate the design.
This research utilizes a truncation-based FP4/FP8 inference approach to streamline hardware design while minimizing area and power usage. The studies referenced in \cite{ref14,ref16,ref17} investigate low-precision floating-point representations, such as FP4 (E3M0, E3M4) and FP8 (E5M2, E4M3), mainly to enhance training efficiency, while~\cite{ref15} uses FP8 to decrease memory area requirements. Meanwhile,~\cite{ref18} introduces a multiple-precision fused multiply--add (FMA) unit that 
\begin{figure}[!t]
\centering
\includegraphics[width=3.4in, height=2.3in]{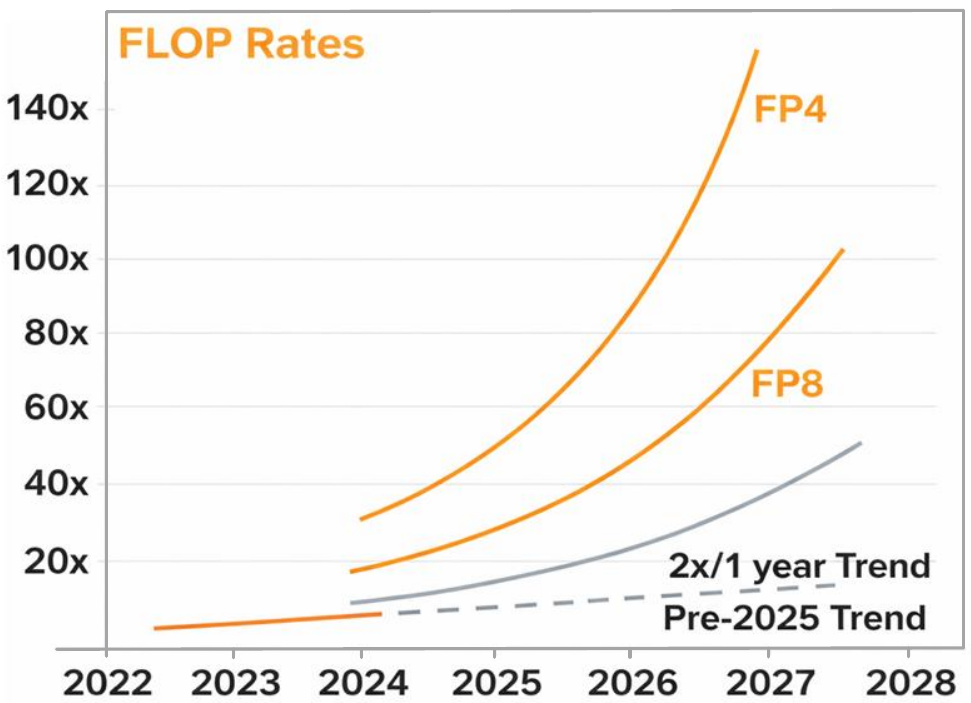}
\caption{ Accelerated GPU performance gains enabled by reduced-precision
numerical formats. Source: AMD\cite{ref21}.}
\label{fig:amd_precision}
\end{figure}
allows for hardware reuse across various floating-point formats for AI tasks, whereas the works cited in~\cite{ref19} examine low-precision numerical formats using different approaches, primarily aiming at edge-AI applications. In contrast,~\cite{ref20} suggests a bit-split combination MAC architecture that simultaneously utilizes both low- and high-precision calculations to optimize deep neural network (DNN) processes. However, these approaches are not optimized for inference-centric workloads, where extremely low-precision computations are essential for improving throughput and energy efficiency. In contrast, this work presents an inference-oriented processing element (PE) incorporating a compact FP4/FP8 multiply--accumulate (MAC) unit \cite{sharma}, specifically designed to reduce computational complexity and hardware overhead.

\begin{figure}[!t]
\centering
\includegraphics[width=3.0in, height=5.5in]{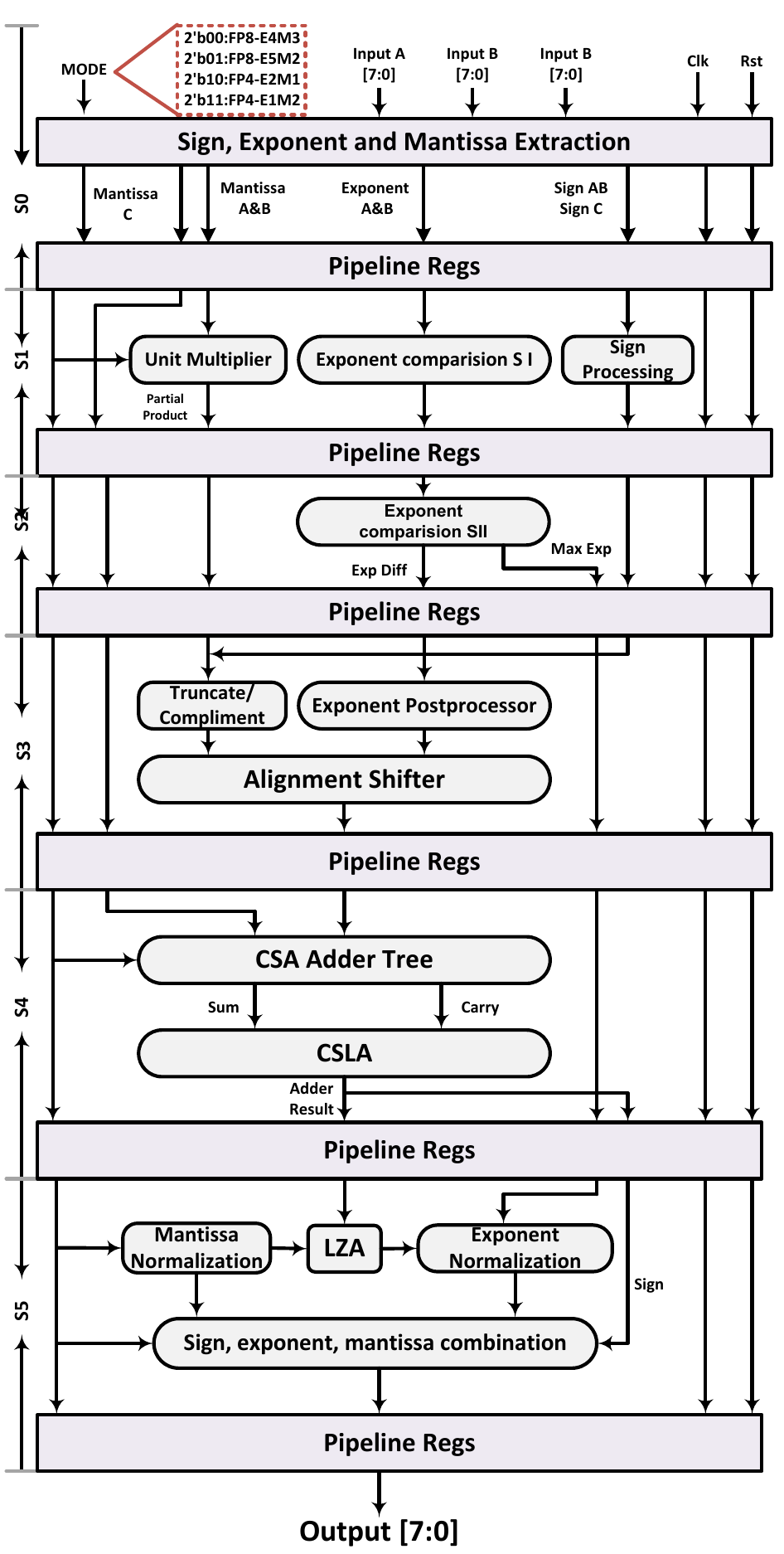}
\caption{Datapath of the proposed fully-pipelined dual-precision PE.}
\label{fig:dataflow}
\end{figure}

Recent developments in GPU performance scaling suggest a significant transformation in how computational throughput is employed for modern AI applications. As demonstrated in Fig.~\ref{fig:amd_precision}, drawing from performance predictions and trends provided by AMD, the performance gains anticipated up to 2025 show a modest increase of about $2\times$ over a span of two years, largely fueled by progress in higher-precision computations. With the swift growth of deep learning model sizes and increasing data volumes, emerging GPU architectures place greater emphasis on low-precision, AI-optimized computations, achieving approximately annual performance doubling starting in 2025. In this framework, reduced-precision formats such as FP8 and FP4 exhibit significantly higher FLOP growth compared to traditional precision formats, highlighting their importance in sustaining performance scaling under stringent power and area constraints. This shift motivates the development of hardware architectures optimized for low-precision floating-point computation, enabling high throughput while preserving numerical accuracy. Consequently, specialized FP4/FP8 arithmetic units and enhanced multiply--accumulate (MAC) architectures serve as key enablers for next-generation AI accelerators.
\begin{figure}[!t]
\centering
\includegraphics[width=3.39in, height=1.3in]{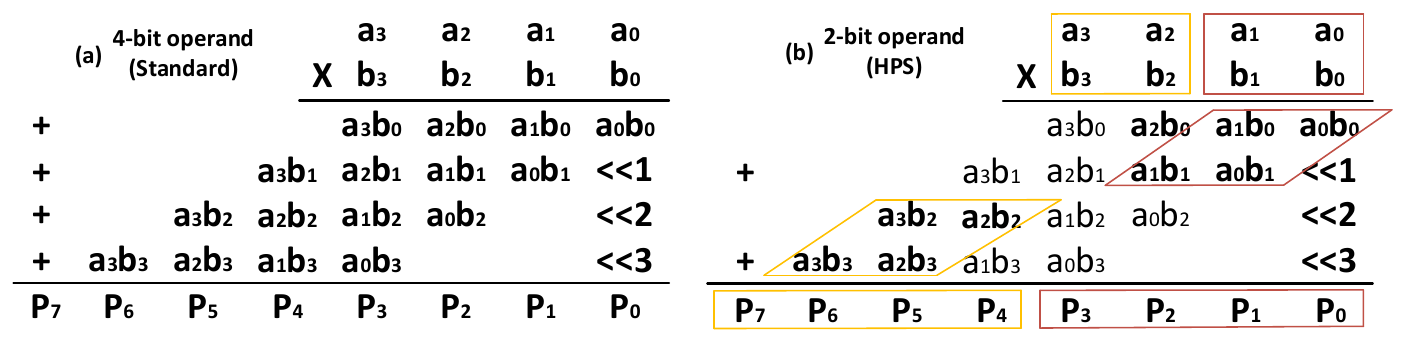}
\caption{Proposed bit-partitioning method and Unit Multiplier: (a) 4-bit operand for FP8; (b) 2-bit operand for FP4}
\label{fig:multiplier}
\end{figure}
The result sign, mantissas are multiplied, and exponents are summed with appropriate bias adjustment, while accumulation represents the summation of multiple products. Lower-precision formats such as FP4 and FP8 follow the same structural principles, as illustrated in Fig.~\ref{fig:dataflow}, but with reduced bit width, offering lower storage requirements, improved bandwidth efficiency, and faster computation compared to FP16 or FP32. Furthermore, integrating this type of accelerator with a RISC-V processor can significantly enhance its applicability in biomedical devices~\cite{Bio-RV}.
\begin{equation}
A = (-1)^{S_a} \times (1.M) \times 2^{E_a - \text{bias}} 
\tag{1}
\label{eq:eq1}
\end{equation}

\begin{equation}
B = (-1)^{S_b} \times (1.M) \times 2^{E_b - \text{bias}} 
\tag{2}
\label{eq:eq2}
\end{equation}

\begin{equation}
A \cdot B =
\sum_{n=0}^{N-1}
(-1)^{S_{an} + S_{bn}}
\left( M_{an} \times M_{bn} \right)
2^{E_{an} + E_{bn} - 2\times \text{bias}}
\tag{3}
\label{eq:eq3}
\end{equation}

\subsection{Bit Partitioning and Unit Multiplier }
To support FP8 precision with a 4-bit unit multiplier, Fig.~\ref{fig:multiplier} illustrates a bit-partitioning approach. For FP4 operation, the 4-bit multiplier is logically divided into two independent 2-bit units, allowing the same hardware to execute either a conventional unsigned $4 \times 4$ multiplication or two parallel unsigned $2 \times 2$ multiplications without additional circuitry. Each operand is split into higher $(a_3, a_2)$ and lower $(a_1, a_0)$ 2-bit segments, which are processed independently using High-Performance Synthesis (HPS). Fig.~\ref{fig:mac_type}(a) shows the standard 4-bit multiplication, while Fig.~\ref{fig:mac_type}(b) demonstrates concurrent dual 2-bit MAC operations. 
Equation~(4) describes the operation of the proposed unit multiplier using a mode-dependent masking function. Each term $(a_i b_j)2^{i+j}$ represents a weighted partial product of the 4-bit mantissas, while the masking function $\delta_m(i,j)$ determines which partial products are activated.
In FP8 mode $(m = 0)$, all partial products are enabled, resulting in a conventional $4 \times 4$ unsigned mantissa multiplication that produces a single 8-bit output. In contrast, in FP4 mode $(m = 1)$, cross-partition partial products are disabled, allowing the same hardware to execute two independent $2 \times 2$ multiplications concurrently. This bit-partitioning framework enables reconfiguration without additional multiplier hardware. By fully utilizing the unit multiplier across precision modes, the design reduces area, power, and critical-path delay while supporting higher operating frequencies. Dynamic precision switching further allows adaptation to varying computational demands, improving overall energy efficiency.
\begin{equation}
P^{(m)} =
\sum_{i=0}^{N}
\sum_{j=0}^{N}
\delta_m(i,j)\,(a_i b_j)\,2^{i+j}
\tag{4}
\label{eq:eq4}
\end{equation}

\subsection{Implementation of variable precision MAC}
\begin{figure}[!t]
\centering
\includegraphics[width=3.4in, height=1.5in]{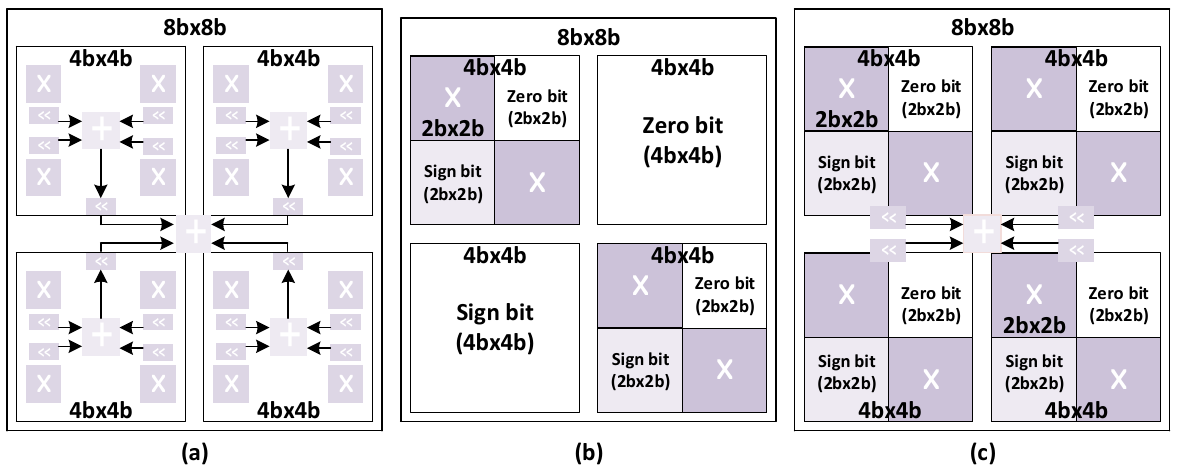}
\caption{Implementation of variable precision MAC. (a) Combination MAC.(b) Split MAC. (c) Bit-Split-and-combination MAC.}
\label{fig:mac_type}
\end{figure}

Fig.~\ref{fig:mac_type} shows three typical architectures for performing variable-precision MAC operations: (a) a combination MAC, (b) a split MAC, and (c) a bit-split-and-combination MAC. The combination MAC depicted in Fig.~\ref{fig:mac_type}(a) employs a fixed full-precision datapath, which leads to inefficient hardware utilization and unnecessary power consumption when operating at lower precisions. In contrast, the split MAC illustrated in Fig.~\ref{fig:mac_type}(b) enables parallel low-precision computations but incurs additional overhead due to shift, alignment, and recombination logic, along with increased control complexity.

\begin{figure}
    \centering
    \includegraphics[width=1\linewidth]{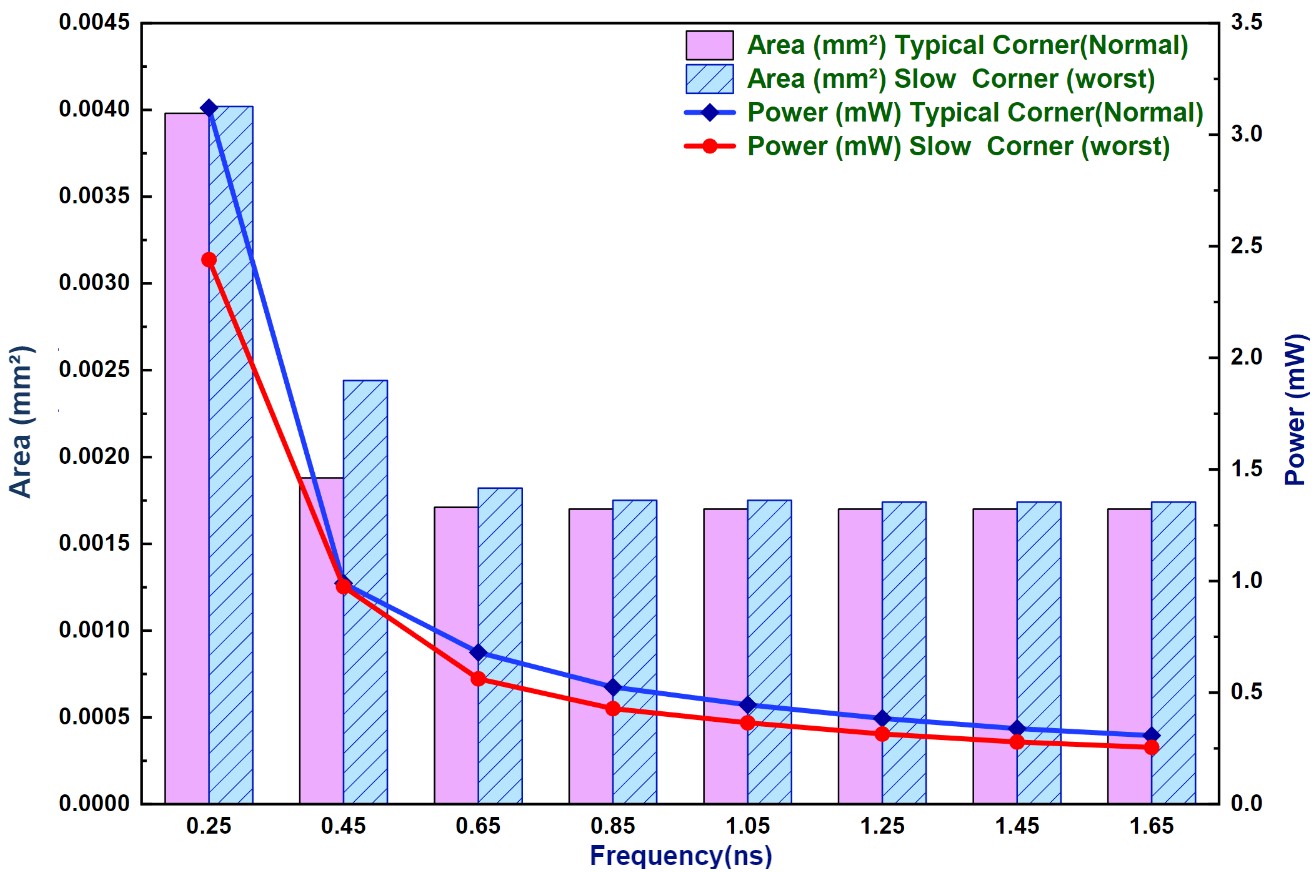}
    \caption{{Area and power variations subject to different clock period constraints: Typical corner ; Slow corner.}}
    \label{fig:area_power}
\end{figure}
To address these limitations, this work adopts the bit-split-and-combination MAC shown in Fig.~\ref{fig:mac_type}(c), which integrates the advantages of both combination and split architectures. Specifically, we introduce a 4-bit split-and-combination MAC that operates in two modes: (i) a conventional $4 \times 4$ multiplication mode and (ii) a split mode that performs two $2 \times 2$ multiplications in parallel. By utilizing 
a shared multiplier array across both modes, the proposed architecture 
improves datapath utilization while minimizing the shift and recombination overhead commonly associated with bit-split MAC designs and avoiding the hardware inefficiencies of split-only configurations.
Furthermore, the ability to exploit increased parallelism at lower precision while maintaining full-precision support enhances energy efficiency, scalable throughput, and silicon-area utilization. Consequently, the proposed bit-split-and-combination MAC is well suited for mixed-precision deep neural network workloads and integrates seamlessly into the target accelerator architecture.

\begin{figure}[!t]
\centering
\includegraphics[width=3.39in, height=2.2in]{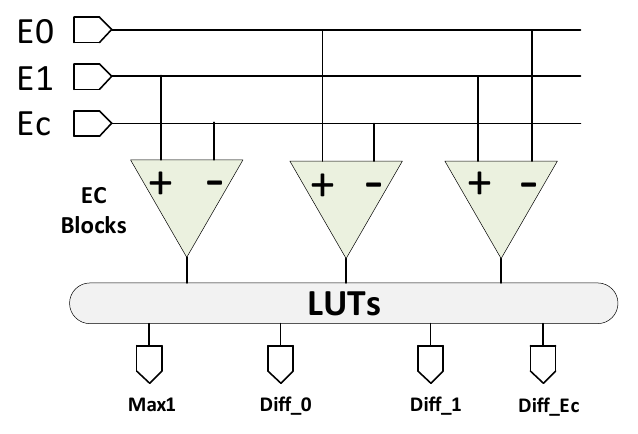}
\caption{Exponent Comparison Using EC+LUT for Mixed-Precision MAC.}
\label{fig:ec_lut}
\end{figure}
\section{Proposed Architecture}
\subsection{Architecture of a 6-stage pipeline}
The dataflow depicted in Fig.~\ref{fig:dataflow} outlines the pipelined structure of the proposed reconfigurable processing element (PE), capable of executing FP8 (E4M3, E5M2) and FP4 (2 x E2M1, 2 x E1M2) formats. Processing begins with the extraction of the sign, exponent, and mantissa fields from each operand. These components 
then follow dedicated paths through the pipeline. The mantissa path incorporates a reconfigurable unit multiplier that supports precision scaling through bit partitioning, enabling either full 8-bit computation or two parallel 4-bit operations for FP4 dual mode.

\textit{S0 (Input Processing):}
To ensure reliable functionality with low-precision formats, the processing
element (PE) employs 8-bit values for the operands $A$, $B$, and $C$.
The selected operating mode (E4M3, E5M2, E2M1, or E1M2) determines the
corresponding bit-decoding method, while formats with fewer than 8 effective
bits apply zero-padding to preserve a uniform input width across all pipeline
stages. In floating-point modes, the operands are decomposed into sign,
exponent, and mantissa fields, with hidden-bit reconstruction applied as
required. Special values, including zeros, subnormal numbers, infinities, and
NaNs, are detected at an early stage to prevent computational errors. In the
dual-FP4 mode, the 8-bit input is partitioned into two independent FP4 values, enabling FP8, FP4, and mixed-precision MAC operations through a shared,
mode-aware datapath.

\textit{S1 (Comparator and Multiplier Array):} 
In the proposed processing element (PE), the sign bits of the operands are combined using an XOR operation to determine the sign of the resulting product. Each 4-bit multiplier block
acts as a partial-product resource, enabling either a single $4 \times 4$ mantissa multiplication or two parallel $2 \times 2$ multiplications through bit partitioning. As a result, one $4 \times 4$ and two $2 \times 2$ multiplications
require only two multiplier blocks, enabling FP8 and dual-FP4 support without expanding the computational array. The PE also integrates a three-input EC+LUT--based exponent comparator to process both operands and the MAC addend, determining the maximum and reference exponents for mantissa alignment while reducing the overall logic depth.
\begin{table}[!t]
\centering
\caption{Comparison of FPGA Resource Utilization}
\label{tab:fpga_comparison}

{\fontsize{7.5}{9}\selectfont
\setlength{\tabcolsep}{6pt}
\renewcommand{\arraystretch}{1.15}
\begin{tabular}{|l|c|c|c|c|c|c|}
\hline
\textbf{Design} &
\begin{tabular}[c]{@{}c@{}}\textbf{FPGA}\\\textbf{Platform}\end{tabular} &
\begin{tabular}[c]{@{}c@{}}\textbf{Freq.}\\(MHz)\end{tabular} &
\begin{tabular}[c]{@{}c@{}}\textbf{Power}\\(mW)\end{tabular} &
\textbf{LUT} &
\textbf{FF} &
\textbf{IO} \\
\hline
TVLSI'25~[5]  & VC707  & 263   & 0.35  & 405  & 116  & -- \\
\hline
TVLSI'22~[6]  & VCU129 & 357   & 0.378 & 8065 & 1072 & 357 \\
\hline
TCASII'24~[7] & VCU129 & 216.5 & 0.296 & 8054 & 1718 & 357 \\
\hline
TCASI'25~[8]  & ZCU102 & 173.2 & 0.061 & 1971 & 1096 & -- \\
\hline
\textbf{This Work} & \textbf{VCU129} & \textbf{180} & \textbf{0.028} & \textbf{106} & \textbf{69} & \textbf{29} \\
\hline
\end{tabular}
}
\end{table}

\begin{table*}[!t]
\caption{Comparison of Multi-Precision MAC/Accelerator Designs}
\centering
\renewcommand{\arraystretch}{1.3}

\begin{tabular}{|c|c|c|c|c|c|c|c|c|c|c|c|c|}
\hline

\multirow{3}{*}{\makecell{\textbf{Design}}} &
\multirow{3}{*}{\makecell{\textbf{Node}\\(nm)}} &
\multirow{3}{*}{\makecell{\textbf{Volt}\\(V)}} &
\multirow{3}{*}{\makecell{\textbf{Freq}\\(MHz)}} &
\multirow{3}{*}{\makecell{\textbf{Area}\\(mm$^2$)}} &
\multirow{3}{*}{\makecell{\textbf{Power}\\(mW)}} &
\multirow{3}{*}{\makecell{\textbf{Precisions}}} &
\multicolumn{3}{c|}{\textbf{Throughput}} &
\multicolumn{3}{c|}{\textbf{Energy Efficiency}} \\
\cline{8-13}

& & & & & & 
& \multicolumn{3}{c|}{\textbf{(GFLOPS)}} 
& \multicolumn{3}{c|}{\textbf{(GFLOPS/W)}} \\
\cline{8-13}

& & & & & & 
& \textbf{FP16} & \textbf{FP8} & \textbf{FP4}
& \textbf{FP16} & \textbf{FP8} & \textbf{FP4} \\
\hline

TVLSI’21 [1] & 22 & 0.8 & 923 & 0.049 & 8.6 & FP64--FP8 
& 12.67 & 25.33 & -- & 1180 & 2950 & -- \\
\hline

GLSVLSI’23 [2] & 28 & 0.8 & 2500 & 0.00987 & 48.4 & FP4--FP8, FP12 
& 20 & 20 & -- & 413 & 413 & -- \\
\hline

ISCAS’24 [3] & 28 & 1.0 & 160 & 1.84 & 67.4 & FXP4/8, FP16, FP32 
& 157 & 157 & -- & 2329 & 2329 & -- \\
\hline

ASPDAC’25 [4] & 28 & 0.9 & 200 & 0.0016 & 0.234 & FP4--FP8, FP12 
& -- & 0.46 & 1.39 & -- & 1948 & 5940 \\
\hline

TVLSI’25 [5] & 28 & 0.9 & 1370 & 0.0491 & 7.34 & FP16, FP8, FP4 
& 2.74 & 2.74 & 2.74 & 373.3 & 373.3 & 373.3 \\
\hline

\textbf{This Work} & 28 & 1.0 & \textbf{1938} & \textbf{0.0039} & \textbf{2.134} & \textbf{FP4--FP8} 
& -- & \textbf{3.88} & \textbf{7.75} & -- & \textbf{1818} & \textbf{3632} \\
\hline

\end{tabular}
\label{tab:comparison}
\end{table*}

\begin{table}[!t]
\centering
\caption{Performance Summary of the Proposed PE Design}
\label{tab:stage_tab}

{\fontsize{7.5}{9}\selectfont
\setlength{\tabcolsep}{6pt}
\renewcommand{\arraystretch}{1.15}
\begin{tabular}{|l|c|c|c|}
\hline
\textbf{Stage} & \textbf{Area ($\mu$m$^2$)} & \textbf{Power (mW)} & \textbf{Delay (ns)} \\
\hline
S0 (Input Decode)      & 582.12  & 0.314  & 0.076  \\
\hline
S1 (Multiplication)   & 1137.31 & 0.613  & 0.148  \\
\hline
S2 (Alignment Shifter) & 370.26  & 0.200  & 0.048  \\
\hline
S3 (Accumulation)      & 928.22  & 0.500  & 0.121  \\
\hline
S4 (Normalization)     & 774.18  & 0.417  & 0.101  \\
\hline
S5 (Output Encode)     & 166.72  & 0.0898 & 0.0216 \\
\hline
\textbf{Total}         & \textbf{3960.00} & \textbf{2.1338} & \textbf{0.5158} \\
\hline
\end{tabular}
}
\end{table}

\textit{S2 (Alignment Shifter)}:
The partial products in Stage~2 are designed to reduce hardware
complexity while maintaining sufficient accuracy for low-precision
AI applications. A truncation-based approach is employed to remove
least significant bits (LSBs), thereby reducing arithmetic complexity,
switching activity, and silicon area. When required, a complement
operation is applied to simplify the accumulation of negative partial
products. The resulting partial products are subsequently aligned
using an alignment shifter based on precomputed exponent differences,
enabling efficient mantissa alignment with reduced hardware cost for
FP4/FP8 deep neural network inference.

\textit{S3 (CSA Adder Tree):}
At this phase, the aligned partial products are initially reduced using a 3:2 carry-save adder (CSA) tree, which combines multiple inputs into sum and carry vectors without immediate carry propagation. This approach significantly shortens the critical path and enables high-throughput pipelined operation. The resulting sum and carry outputs are subsequently combined using a carry-select adder (CSLA), which accelerates the final addition by precomputing results for different carry-in conditions, thereby reducing carry-propagation delay. This two-level accumulation strategy achieves an effective trade-off between speed and hardware efficiency, making it particularly well suited for low-precision, high-throughput multiply–accumulate (MAC) operations in deep neural network inference.

\textit{S4 (LZA and Normalizer/Truncation):}
S4 includes a leading-zero anticipator (LZA) that counts the number of leading zeros in the mantissa of the MAC result. This information is then used to normalize both the exponent and the mantissa such that the final output conforms to the FP4 or FP8 format. During earlier pipeline stages, exponent alignment may cause the loss of lower-order bits, and truncation can slightly affect the accumulated value. The normalization process mitigates this effect by ensuring that the mantissa is properly scaled and aligned with its corresponding exponent, thereby maintaining the deviation from the reference model within an acceptable range without relying on rounding techniques.

\textit{S5 (Output processing/A F):}
After normalization, both the mantissa and exponent are adjusted to generate a normalized floating-point result, formatted according to either the FP4 or FP8 specification. The resulting value is then processed by a rectified linear unit (ReLU) activation function, which is preferred in hardware implementations due to its low complexity and reliance on simple sign-bit comparisons. The inclusion of ReLU introduces essential nonlinearity into the computation, enabling the accelerator to model complex decision boundaries beyond linear transformations. By suppressing negative values, the ReLU operation improves inference robustness while incurring minimal latency and negligible critical-path overhead.

\subsection{EC+LUT-Based Exponent Comparator}
In mixed-precision floating-point MAC units, exponent comparison identifies the maximum exponent and determines mantissa alignment shifts. As parallelism increases, enumeration-based comparators become inefficient due to rapidly growing comparator and interconnect complexity. Hierarchical pairing reduces duplication but increases comparison depth, while cascaded exponent comparison (CEC) and pipeline enumeration comparison (PEC) lower hardware cost at the expense of long multi-stage critical paths caused by sequential exponent accumulation~\cite{ref6,ref7}.
 
Fig.~\ref{fig:ec_lut} illustrates a three-input EC+LUT--based exponent comparator integrated into the processing element (PE) for FP4/FP8 formats. Parallel EC blocks compute exponent differences, which are processed by a shared lookup table (LUT) to determine the maximum exponent and generate mantissa alignment offsets in a single decode stage. By replacing complex comparison logic with a fixed-latency LUT, the design reduces logic depth and critical-path delay, making it well suited for low-precision, high-throughput deep neural network inference.

\section{Experimental Results}
\subsection{Performance Evolution}
To ensure the reliability of the proposed processing element (PE), the design is implemented using the TSMC 28-nm technology and synthesized under both typical (1.00~V, 25~$^\circ$C) and slow (0.90~V, 110~$^\circ$C) process corners using Cadence Genus tool. The functional correctness is confirmed using a specialized testbench, and synthesis is carried out to assess the performance, power, and area (PPA) attributes of the design.
To determine the optimal PPA operating point, the PE is synthesized across clock period constraints $[0.3:0.1:1.7]$~ns. The corresponding variations in area and power under both typical and slow corners are shown in Fig.~\ref{fig:area_power}. As the clock period is relaxed, area and power reduce significantly under tight timing constraints and gradually stabilize beyond moderate clock periods, indicating convergence after timing optimization.
The synthesis results show that the PE achieves timing closure at approximately 0.65~ns under the typical corner and 0.85~ns under the slow corner. Beyond these points, further clock relaxation offers minimal PPA improvement. Consequently, the typical-corner outcomes associated with a 0.85~ns clock period are selected as the representative operating point, striking a balanced compromise among performance, power consumption, and silicon area.

\subsection{Performance Comparison}
The presented work achieves 7.75~GFLOPS (FP4) and 3.88~GFLOPS (FP8) at an operating frequency of 1.938~GHz, significantly exceeding the results reported in ASPDAC’25, which achieved 1.39~GFLOPS (FP4) and 0.46~GFLOPS (FP8). This corresponds to approximately $5.6\times$ higher FP4 performance and $8.4\times$ higher FP8 performance, highlighting the benefits of higher operating frequency and an optimized datapath. 
In terms of energy efficiency, the proposed architecture achieves 3632~GFLOPS/W (FP4) and 1818~GFLOPS/W (FP8), surpassing TVLSI’25 (373~GFLOPS/W) and GLSVLSI’23 (413~GFLOPS/W). Although ASPDAC’25 reports higher FP4 efficiency (5940~GFLOPS/W), the proposed work delivers substantially higher absolute throughput, providing a better performance--power trade-off.
Compared with larger accelerators such as ISCAS’24 and FPnew, the proposed design targets MAC-level efficiency with a very small area (0.0039~mm$^2$), making it suitable for scalable PE arrays and edge-AI systems.

\subsection{FPGA Implementation}
The FPGA implementation results in Table~\ref{tab:stage_tab} demonstrate substantial improvements in resource utilization and power efficiency for low-precision operation compared with prior designs. Compared with TCAS-II'24\cite{ref6}, the proposed architecture reduces LUT usage by 98.7\% and flip-flop count by 98.2\%. Relative to TVLSI'25\cite{ref5}, the design achieves reductions of 73.8\% in LUTs, 40.5\% in flip-flops, and 92.0\% in power consumption for low-precision computation. When compared with TVLSI'22\cite{ref7}, LUT usage is reduced by 98.6\%, while relative to TCAS-I'25\cite{ref8}, LUTs, flip-flops, and power are reduced by 94.6\%, 93.7\%, and 54.1\%, respectively. These results confirm the efficiency of the proposed design for low-precision, resource-constrained FPGA implementations.

\section{Conclusion and Future Work}
This work presents a processing element capable of dual precision, specifically designed for low-bit floating-point multiply--accumulate operations, supporting both FP4 and FP8 formats. Using a 4-bit bit-partitioning multiplier technique, the multiplier array efficiently handles these formats, improving performance while reducing unnecessary computations. The design is versatile enough to support FP4 and FP8 formats, allowing for precision adjustments that align with the requirements of modern high-performance computing and artificial intelligence applications. Future work will broaden the scope of the proposed processing element (PE) to encompass additional numerical formats and low-bit integer representations for both training and inference. The PE will be extended to array-level architectures, such as systolic arrays and SIMD-style designs, RISC V and integrated into a complete AI accelerator with optimized memory hierarchies and dataflow strategies to evaluate system-level throughput, scalability, and energy efficiency for deep neural network applications.

\bibliographystyle{ieeetr}
\bibliography{bib}

\end{document}